# A modified Michelson interferometer type Raman laser system for atom interferometers

Ningfang Song,[1,] Xiangxiang Lu,[1,a)] Wei Li,[1] Yang Li,[1] Yingying Wang,[1] Jixun Liu,[1] Xiaobin Xu,[1] and Xiong Pan[1]

[1]*Institute of Opto-electronics Technology, Beihang University, Beijing 100191, China*

We have developed a modified Michelson interferometer type Raman laser system to manipulate cold $^{87}$Rb atoms to interfere. A frequency-modulated continuous-wave technique was introduced to determine the optical path difference, thus compensating it to zero to minimize the effects of common mode noise. The beat signal's linewidth (full width at half maximum) of the laser system at 6.834 GHz was measured to ~1Hz limited by the resolution bandwidth of the spectral analyzer. The measured rms phase variance of the laser system's phase noise at 166 MHz was 0.015 rad$^2$, mainly restricted by our poor performance radio frequency microwave source. With modest improvements, we plan to apply this laser system to form an atom interferometer for acceleration and rotation measurements.

Atom gyroscopes based on neutral atom interferometry are theoretically ~$10^{10}$ more sensitive to accumulated phase shifts induced by inertial effects,[1] say rotation[2,3] or acceleration,[4,5] compared to their optical counterparts such as fiber-optic and ring laser gyroscopes. One of the key elements to construct such a practical atom interferometer is the beam splitters and the mirrors used to manipulate cold atoms.[6] Since its first proof of principle,[7] stimulated Raman transitions has become one of the most powerful tools to split the atomic wave packets. This kind of two photon transitions allowed the rapid development of highly sensitivity inertial sensors,[8,9] whose performances are now able to compete with some state of the art scientific instruments.[5,10]

To obtain a high signal-to-noise (SNR) ratio and good fringe pattern contrast of interferometer outputs, high performance Raman laser systems, with fixed frequency and phase difference, are favorable and need to be prepared to achieve the desired sensitivity. There are two main techniques to produce Raman lasers. The first method utilizes an optical phase lock loop (OPLL) technique[11-13] to phase-lock two different external cavity diode lasers (ECDLs), the phase noise is determined by the radio frequency (rf) microwave source and the residual noise of the OPLL regardless of vibrations.[14] The latter scheme generates the Raman beams from a common diode laser via frequency modulation processes,[15-17] and the phase noise can be divided into two parts, one is the common mode phase noise of the master laser and the other is the rf source noise without considering the effects of vibrations.[18]

In this paper, we have developed a Raman laser system to manipulate $^{87}$Rb atoms to interfere, thus sensing rotation and acceleration as inertial sensors.[19] The modified Michelson interferometer (MMI) configuration offers great convenience, as is the case in our system, when tuning the optical path difference (OPD) between Raman beams is essential. To eliminate the effects of phase noise on the performance of Raman lasers, a frequency-modulated continuous-wave (FMCW) method[20] is introduced for extracting and compensating the OPD to set the final OPD to zero. With the heterodyne technique,[21] a spectrum analyzer (SA) is used to measure the beat signal's full-width at half maximum (FWHM) linewidth, which is less than 1 Hz. The phase noise spectral density of the Raman laser system is also measured with a signal source analyzer (SSA). The obtained result, -20 dBc/Hz @ 1 Hz at 166 MHz, shows that the laser system's phase noise is mainly limited by the low performance rf source and additional unwanted vibrations. With a better rf source and vibration isolation measures, this laser sys-

a) luxiangxiang@buaa.edu.cn

tem can be at the basis of a highly sensitive cold $^{87}$Rb atom interferometer.

The experimental setup is shown in Fig. 1. The laser system mainly consists of an ECDL owning a linewidth of 100 kHz and a modified Michelson interferometer. When operating the laser at a temperature of 20.3 °C and a current of 236 mA, the output power, before and after a 60 dB optical isolator (OI), is 92 mW and 81 mW, respectively. The mode-hop free tuning range (MHFTR) of the laser is ~ 20 GHz, making it quite easy to observe the complete D$_2$ line spectroscopy of $^{87}$Rb atoms. A maximum MHFTR of 53 GHz can be obtained with a laser current of 212 mA.

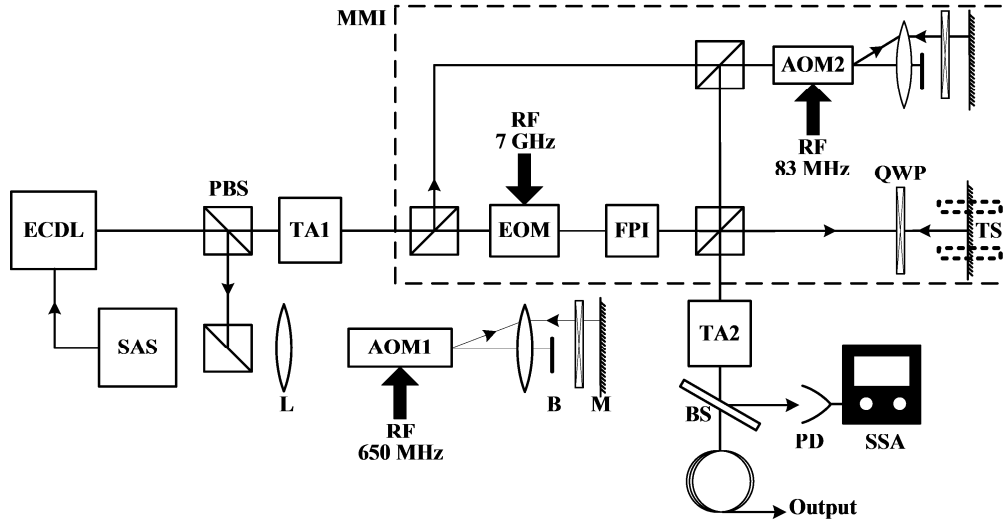

Fig. 1. Schematic of the experimental setup. L: Lens; B: Block; M: mirror; EOM: Electro-optic modulator; BS: Beam splitter; PD: Photo detector; TS: Translation stage; QWP: Quarter wave plate. Other acronyms can be found in the context elsewhere. Note that optical isolators, half wave plates and other essential optical opponents are omitted here for simplicity.

To keep the output optical frequency as stable as possible, a fraction of optical power is reflected by a polarizing beam splitter cube (PBS), then locked to a transition line of $^{87}$Rb via saturated absorption spectroscopy (SAS) and the error signal was fed back to control the piezoelectric transducer (PZT) voltage and the drive current. In order to reject spontaneous emissions during the Raman transition process, a red frequency detuning of 1.2 GHz to 1.4 GHz can be realized by double passing the light through a 650 MHz frequency shifter. Due to the extremely small optical aperture, a lens with 100 mm focal length is inserted to focus the beam into the required size. A block is placed behind the lens to prevent the zero order light reflecting back to the acousto-optic modulator (AOM). Thanks to this double pass configuration, the light direction will not change much while the frequency of the AOM is slightly tuned from its center. The sensitivity of the alignment to the input frequency setting of the AOM is thus greatly reduced by using this cat's eye configuration. About 20 mW is sent into a tapered amplifier (TA) and as much as 880 mW output power can be obtained with a working current of 1.65 A at a temperature of 20.8 °C, providing enough optical power for frequency modulation usage.

To generate the two Raman beams with orthogonal polarization and fixed frequency and phase difference at 6.834 GHz, a frequency modulation technique is applied in this laser system. The light amplified by TA1 is first divided by a PBS combined with a half wave plate, then sent to AOM2 and EOM with moderate power, respectively. The light through the EOM is frequency shifted by 7 GHz with a home-made microwave reference. To choose the desired sideband, a Fabry-Perot interferometer with a free spectral range (FSR) of 2.5 GHz is used to serve as an optical filter. The other light is shifted 166 MHz by double passing the light through an 83 MHz AOM2. To reduce the laser phase noise and other types of common mode noise, a modified Michelson interferometer configuration is constructed to compensate the OPD between the Raman beams. A mirror is placed on a translation stage in one arm, so that the OPD between the AOM2 arm and the EOM arm can be adjusted using the FMCW method, with an adjustment precision better than 50 nm. At the output port of the MMI system, a second tapered amplifier (TA2) is used to amplify the optical power. About 1 mW of the beat signal is detected by a high bandwidth photo detector and the electronic signal is sent to a SSA (SA) for phase noise (linewidth) measurement. The rest of the Raman beams, the optical power can be up to 1.78 W when TA2 is operated with a current of 2.58 A at a temperature of 20.8 °C, are coupled into a polarization maintaining fiber for driving $^{87}$Rb atoms undergo stimulated Raman transitions.

This MMI configuration design has several advantages over conventional Raman laser system schemes. Firstly, the overall laser system originated from the same seed laser source, making it feasible to reduce the phase noise in principle. Secondly, light in the two arms both went through a same sequence in OPD, namely two reflections and two transmissions, so they are optically reciprocal except the polarization, which means common noise can be effectively rejected. Finally, the OPD can be precisely resolved with the FMCW method and then compensated with a translation stage at will, minimizing the effects of common mode noise.

The FMCW method was originally studied in radar in 1950s and the principle of this technique can be found in many references,[22-25] so we will introduce this method briefly and present our results with this technique. According to Ref. 20, the beat signal in a triangular wave interference can be expressed by

$$I(\tau,t) = I_0[1 + \eta \cos(2\pi\alpha\tau t + \omega_0\tau)] \qquad (1)$$

where $I_0$ is the average intensity of the beat signal, $\eta$ the contrast of the fringe pattern, $\alpha$ the chirp rate of frequency modulation, $\omega_0$ the angular frequency at the center of the rising period, $\tau$ the delay time. By measuring the frequency of the beat signal, the delay time $\tau$ and the OPD of Raman beams can be obtained by relating to Equ. (1). The interference waveform produced by two coherent light whose frequency modulated by a triangular wave is shown in Fig. 2, where $T_m$, ~100 ms in our case, is the period of the triangular wave. Typically, the phase jumps abruptly to the opposite direction when the PZT controlled external cavity changes its direction.

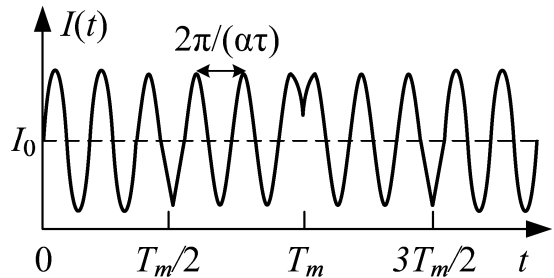

Fig. 2. Beat signal generated by two triangular waves modulated with the FMCW method.

By modulating the laser frequency with a periodic triangular wave, we obtain the beat signal with a photo detector and the measured beat signal is displayed in Fig. 3. In our measurement, the period of triangular wave is set to 100 ms and the

chirp rate is set to 125 MHz/ms. The OPD can then be deduced by the following equation [24]

$$OPD = c\tau = c/(\alpha T) \quad (2)$$

where $c$ is the speed of light in vacuum space, $T$ the period of the beat signal. After performing the above procedures, the period of the beat signal in Fig. 3 is determined to 10.15 ms, and the corresponding OPD is calculated as 23.6 cm, which is in good agreement with the expected 23 cm. The translation stage provides a precision better than 50 nm, so it is quite convenient to change the OPD by driving the translation stage. We have successfully adjusted the OPD to zero, while in this case the photo detector outputs a dc voltage corresponding to $I_0$. In our measurement, we find that the period of the beat signal gets smaller as the triangular wave is rising (vice versa), a typical set of values are 10.53 ms, 10.16 ms and 9.77 ms, respectively. A possible explanation is due to the nonlinearity of the frequency scanning, namely when the triangular wave is applied on the PZT, the frequency range at first is large. However, as the external cavity is close to its end, this frequency range change becomes nonlinear.

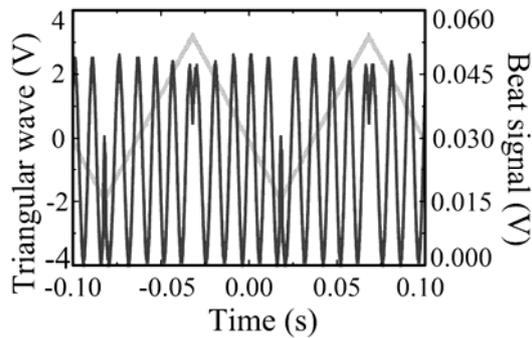

Fig. 3. The light grey one is the triangular wave, while the black one is the beat signal produced. The dc voltage beat signal corresponding to the zero OPD is not presented here.

Due to the poor performance of the phase noise of AOM2 rf source, we failed to measure the laser system's phase noise at 6.834 GHz, since the phase noise of the EOM rf at 6.834 GHz is much lower than that of the AOM2 rf at 166 MHz, -30 dBc/Hz@1Hz and -20 dBc/Hz@1Hz, respectively. However, we performed the phase noise measurement of the Raman lasers at 166 MHz, as can be seen in Fig. 4, to assess the degradation of vibration on the laser system's phase noise. The lower trace is the phase noise power spectral density (PSD) of the 83 MHz rf used to drive AOM2, the middle trace transposed from the 83 MHz to 166 MHz assuming no degradation and the upper trace the phase noise PSD of the laser system. In frequencies from 1 Hz to 10 Hz and from 3 kHz to higher frequency, the phase noise of Raman lasers is in good agreement with that of 166 MHz rf. The slight mismatch is due to the measurement uncertainty of SSA. Between 10 Hz to 3 kHz, the phase noise is much higher than that of 166 MHz rf, mainly due to the vibration effects. The overall rms phase variance of the laser system at 166 MHz is then calculated to be 0.015 rad$^2$.

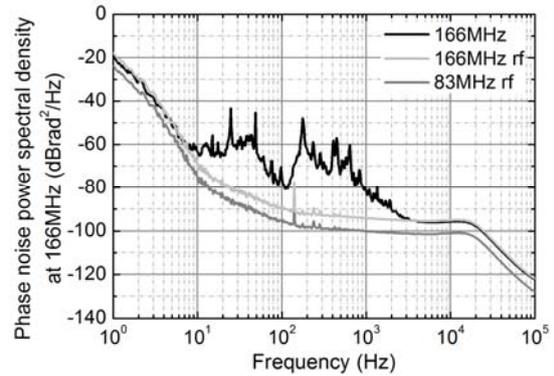

Fig. 4. Power spectral density of the phase noise of the Raman laser system signal at 166 MHz (in black), of the rf signal at 83 MHz (in grey) and transposed at 166 MHz assuming no degradation (in light grey).

With the heterodyne method,[25] the beat signal's linewidth is measured with a SA. Fig. 5 shows the detected spectrum,

which reveals a linewidth better than 1 Hz for the MMI laser system with our self-made rf source used to drive the AOMs and the EOM. The measured ~ 1 Hz linewidth is mainly limited by the resolution bandwidth (RBW) of the SA.

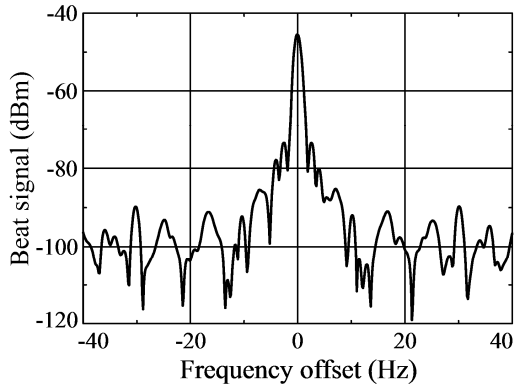

Fig. 5. Power spectrum of the beat signal at $f_0$ = 6.834 GHz between the Raman beams. The spectrum analyzer's resolution bandwidth is set to 1 Hz. The full trace is swept in 1.9 s. The measured linewidth of ~1 Hz is mainly limited by the RBW of the SA.

To conclude, we have developed a modified Michelson interferometer type Raman laser system for cold $^{87}$Rb atom interferometers. A frequency-modulated continuous-wave method was applied to determine the OPD between the Raman beams. We have successfully tuned the OPD to zero with a mirror placed on a translation stage. The linewidth of the beat signal at 6.834 GHz was measured with a SA. The measured linewidth of ~1 Hz is mainly limited by the finite RBW of the instrument. The phase noise measurement of the Raman laser system at 166 MHz is performed and the obtained result shows a total rms phase variance of 0.015 rad$^2$, mainly due to the poor performance of the 166 MHz rf source. Improvements are under way to further lower the phase noise and increase the optical power instability, providing a powerful tool for atom gyroscopes as well as other types of atom interferometer based instruments.

The authors would like to thank Pingwei Lin for fruitful discussions and Shifeng Yang for his contribution to the construction of the Raman laser system.

[1] J. F. Clauser, Phys. B **151**, 262-272 (1988).
[2] T. L. Gustavson, P. Bouyer, and M. A. Kasevich, Phys. Rev. Lett. **78**, 11 (1997).
[3] T. Müller, M. Gilowski, M. Zaiser, P. Berg, Ch. Schubert, T. Wendrich, W. Ertmer, and E. M. Rasel, Eur. Phys. J. D **53**, 273 (2009).
[4] A. Peters, K. Y. Chung, S. Chu, Nature (London) **400**, 849 (1999).
[5] Y. Bidel, O. Carraz, R. Charrière, M. Cadoret, N. Zahzam, and A. Bresson, arXiv: 1302.1518v1, (2013).
[6] P. R. Berman, Academic Press, New York, (1997).
[7] M. Kasevich and S. Chu, Phys. Rev. Lett. **67**, 2 (1991).
[8] M. J. Snadden, J. M. McGuirk, P. Bouyer, K. G. Haritos, and M. A. Kasevich, Phys. Rev. Lett. **81**, 5 (1998).
[9] B. Canuel, F. Leduc, D. Holleville, A. Gauguet, J. Fils, A. Virdis, A.Clairon, N. Dimarcq, Ch. J. Bordé, A. Landragin, P. Bouyer, Phys. Rev. Lett. **97**, 010402 (2006).
[10] K. U. Schreiber, A. Velikoseltsev, M. Rothacher, T. Klugel, G. E. Stedman, D. L. Wiltshire, J. Geophys. Res. **109**, B06405 (2004).
[11] L. Cacciapuoti, M. de Angelis, M. Fattori, G. Lamporesi, T. Petelski, M. Prevedelli, J. Stuhler, and G. M. Tino, Rev. Sci. Instrum. **76**, 053111 (2005).
[12] H. Müller, S. Chiow, Q. Long, and S. Chu, Opt. Lett. **31**, 2 (2006).
[13] M. Nagel, K. Möhle, K. Döringshoff, E. V. Kovalchuck, and A. Peters, Appl. Phys. B **102**, 11 (2011).
[14] P. Cheinet, B. Canuel, F. Pereira Dos Santos, A. Gauguet, F. Yver-Leduc, and A. Landragin, IEEE Trans. Instrum. Meas. **57**, 1141 (2008).
[15] P. Bouyer, T. L. Gustavson, K. G. Haritos, and M. A. Kasevich, Opt. Lett. **21**, 18 (1996).
[16] V. Ménoret, R. Geiger, G. Stern, N. Zahzam, B. Battelier, A. Bresson, A. Landragin, P. Bouyer, Opt. Lett. **36**, 21 (2011).
[17] A. Bonnin, N. Zahzam, Y. Bidel, and A. Bresson, Phys. Rev. A **88**, 043615 (2013).
[18] J. Le Gouët, T. E. Mehlstäubler, J. Kim, S. Merlet, A. Clairon, A. Landragin, F. P. Dos Santos, Appl. Phys. B **92**, 133 (2008).
[19] B. Dubetsky and M. A. Kasevich, Phys. Rev. A **74**, 023615, (2006).
[20] J. Zheng, Springer, New York (2005).
[21] T. Okoshi, K. Kikuchi and A. Nakayama, Electron. Lett. **16**, 630 (1980)
[22] A. J. Hymans and J. Lait, Proc. IEEE **107**-B, 365 (1960).
[23] M. I. Skolnik, Mcgraw-Hill, New York (1962).
[24] R. Schneider, P. Thurmel, M. Stockmann, Opt. Eng. **40**, 33(2001).
[25] J. Zheng, Appl. Opt. **43**, 4189 (2004).